\documentclass[11pt,twoside]{article}

\usepackage{asp2006}
\usepackage{graphicx}

\begin{document}
\title{The Realm of the First Quasars in the Universe: \\ the X-ray View}
%%%%%%%%%%%%%%%%%%%%%%%%%%%%%%%%%%%%%%%%%%%%%%%%%%%%%%%%%%%%%%%%%%%%%%%%%%%%%%%%%%%%%%%%%%%%%%%
%\author{C. Vignali}
%\affil{Dipartimento di Astronomia, Universit\`a degli Studi di Bologna, Italy}
%\author{W.N. Brandt, O. Shemmer, A. Steffen, D.P. Schneider}
%\affil{Department of Astronomy \& Astrophysics, University Park, USA}
%\author{S. Kaspi}
%\affil{Tel Aviv University, Israel; Physics Department, Technion, Haifa, Israel}
%%%%%%%%%%%%%%%%%%%%%%%%%%%%%%%%%%%%%%%%%%%%%%%%%%%%%%%%%%%%%%%%%%%%%%%%%%%%%%%%%%%%%%%%%%%%%%%
\author{C. Vignali (1), W.N. Brandt (2), O. Shemmer (2), A. Steffen (2), D.P. Schneider (2), 
S. Kaspi (3,4)}
\affil{(1) Dipartimento di Astronomia, Universit\`a degli Studi di Bologna, Italy; 
%(2) INAF -- Osservatorio Astronomico di Bologna, Italy; 
(2) Department of Astronomy \& Astrophysics, Pennsylvania State University, University Park, USA; 
(3) Wise Observatory, Tel Aviv University, Israel; 
(4) Physics Department, Technion, Haifa, Israel.}
%%%%%%%%%%%%%%%%%%%%%%%%%%%%%%%%%%%%%%%%%%%%%%%%%%%%%%%%%%%%%%%%%%%%%%%%%%%%%%%%%%%%%%%%%%%%%%%

\begin{abstract}
%%%%%%
%Quasars at $z>4$ provide direct information on the first massive structures to form in the 
%Universe. 
%Here we summarize the X-ray studies of the highest redshift quasars, focusing on the results 
%obtained with \chandra\ and \xmm. Overall, the X-ray and broad-band properties of high-redshift 
%quasars and local quasars are similar, suggesting that the small-scale X-ray emission regions 
%of AGN are insensitive to the dramatic changes occurring at $z$$\approx$0--6. 
%%%%%%
We review the X-ray studies of the highest redshift quasars, focusing on the results 
obtained with {\it Chandra} and XMM-{\it Newton}. 
Overall, the X-ray and broad-band properties of $z>4$ quasars and local quasars are similar, 
suggesting that the small-scale X-ray emission regions 
of AGN are insensitive to the significant changes occurring at $z$$\approx$0--6. 
\end{abstract}

\section{Introduction}
%%%%%%
%X-ray follow-up observations 
%have brought the number of X-ray detections from 6 in 2000 (Kaspi, Brandt, \& Schneider 2000; 
%see Fig.~1, left panel) to more than 100 today (Fig.~1, right panel), 
%and have allowed probing of the inner region of AGN when the Universe was less than 1~Gyr old. 
%%%%%%
In recent years, optical surveys (e.g, the Sloan Digital Sky Survey 
and the Digital Palomar Sky Survey) have discovered a 
large number ($\approx$~1000) of quasars at $z>4$. 
From the pioneering study of Kaspi et al. (2000; see Fig.~1a), 
the number of X-ray detected AGN at $z>4$ has increased to more than 110 
(Fig.~1b), mostly thanks to exploratory observations with {\it Chandra} 
(e.g., Vignali et al. 2001, 2005; Brandt et al. 2002; Bassett et al. 2004; Lopez et al. 2006; 
Shemmer et al. 2006a) and longer exposures with XMM-{\it Newton} (e.g., 
%Brandt et al. 2001; 
%Ferrero \& Brinkmann 2003; Farrah et al. 2004; Grupe et al. 2004, 2006; 
Shemmer et al. 2005). 
At the very faint X-ray fluxes, X-ray surveys have provided detection of several $z>4$ AGN and 
quasars (e.g., Schneider et al. 1998; 
Silverman et al. 2002; Vignali et al. 2002). 
%Silverman 2005; Castander et al. 2003; Treister et al. 2004; Steffen et al. 2004; 
%Eckart et al. 2006; Fontanot et al. 2006). 
%
%%%%%%%%%%%%%%%%%%%%%%%%%%%%%%%%%%%%%%%%%%%%%%%%%%%%%5
% Figure 1: 2000-2006: Status of the art 
\begin{figure}
\includegraphics[angle=0,width=0.45\textwidth]{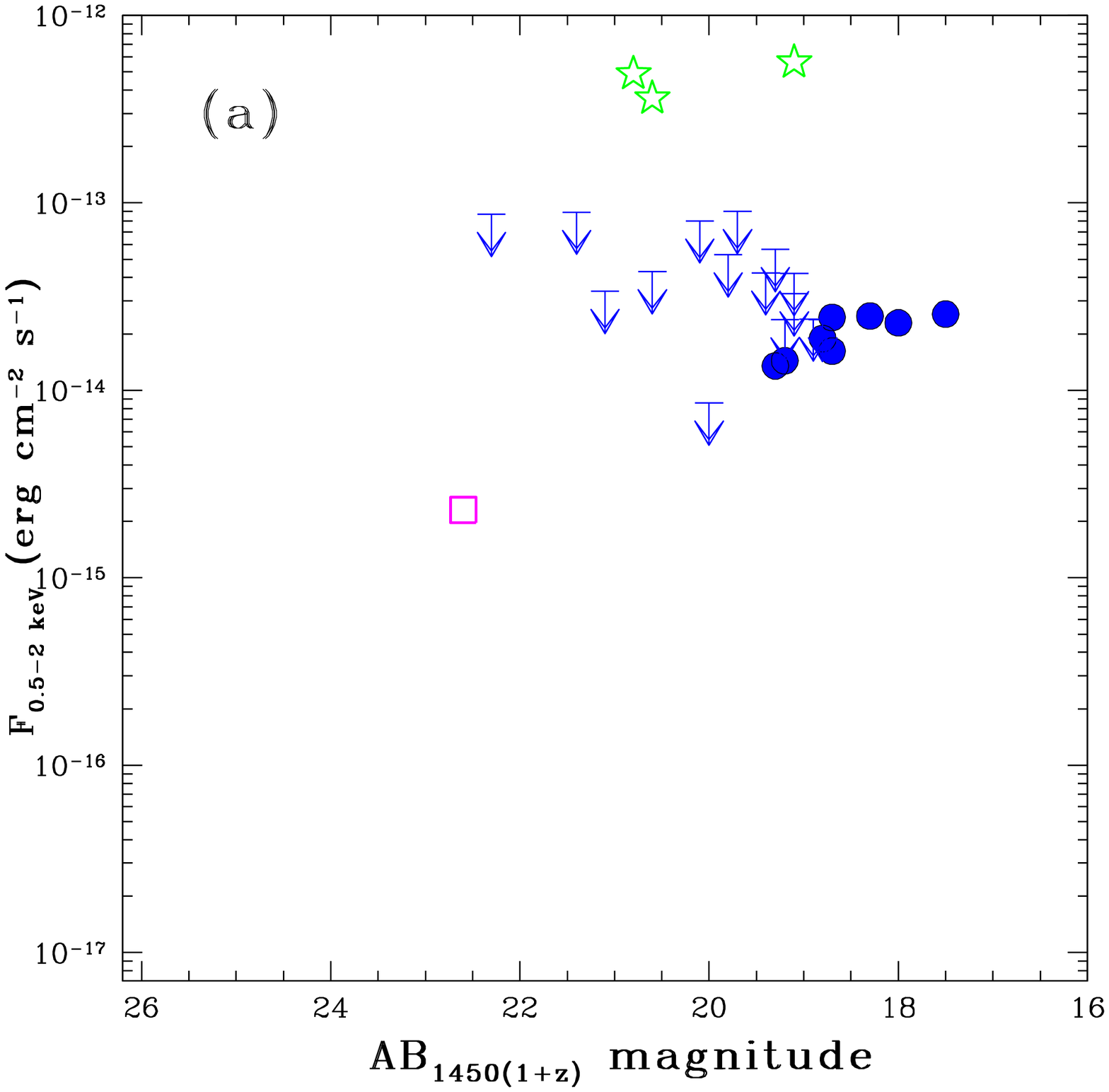}
\hfill
\includegraphics[angle=0,width=0.45\textwidth]{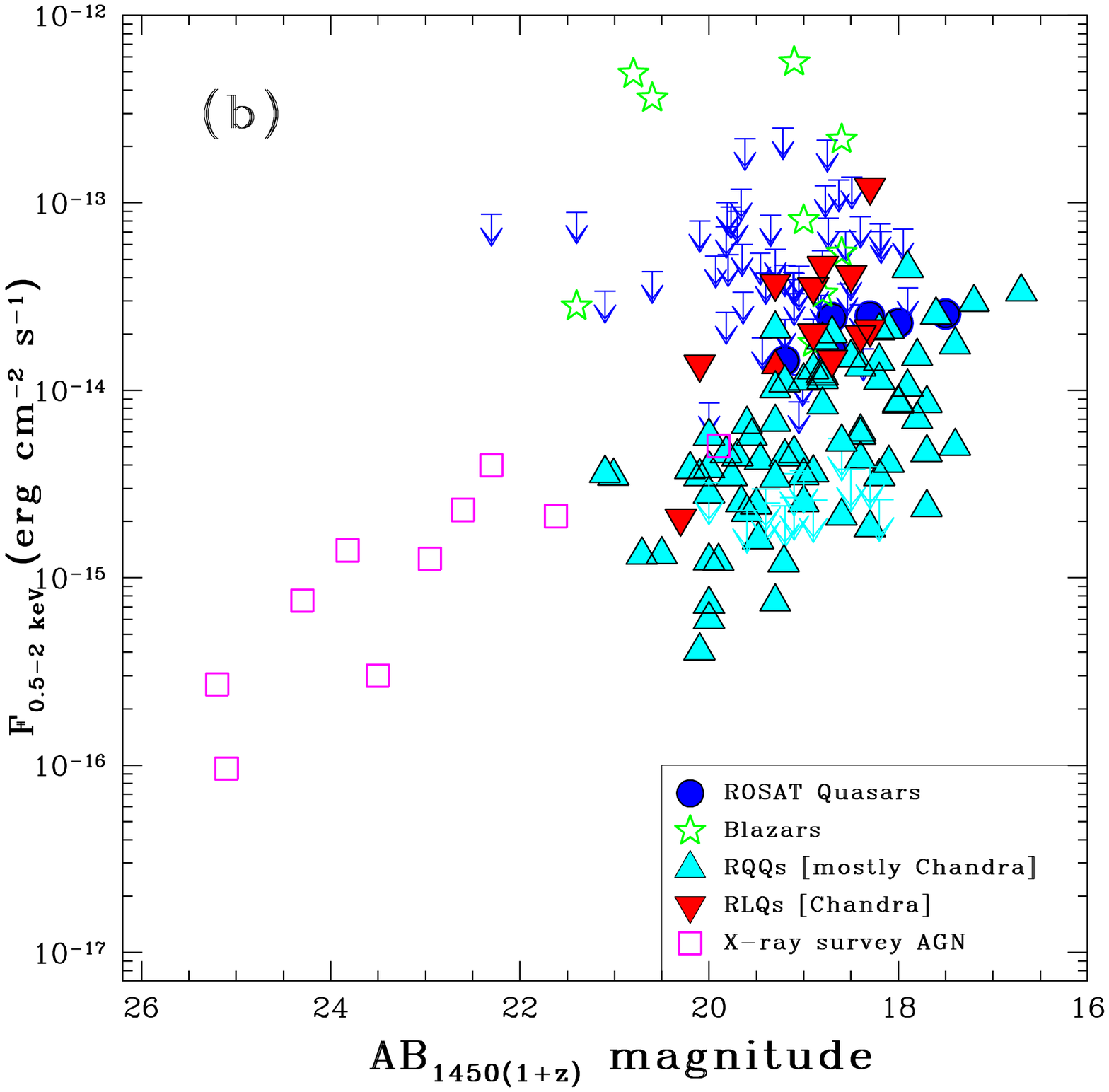}
\caption{Observed-frame, Galactic absorption-corrected \hbox{0.5--2~keV} 
flux versus $AB_{1450(1+z)}$ magnitude for $z>4$ AGN and quasars. 
{\bf (a)} The situation after the Kaspi et al. (2000) work using {\it ROSAT} data; 
{\bf (b)} the updated census of X-ray observations of $z>4$ AGN, including the results 
from moderate-depth and ultra-deep X-ray surveys. 
%See http://www.astro.psu.edu/users/niel/papers/highz-xray-detected.dat for an updated list of 
%X-ray detected AGN at high redshift. 
}
\label{fxab_compendium}
\end{figure}
%%%%%%%%%%%%%%%%%%%%%%%%%%%%%%%%%%%%%%%%%%%%%%%%%%%%%5
%
Here we provide a summary of some of the main recent results: 
\smallskip\par\noindent$\bullet$ 
X-ray emission is a universal property of AGN. %at the highest redshifts. 
The X-ray properties of high-redshift AGN and quasars (derived from either 
stacked or individual X-ray spectra) are similar to those of local quasars, with no evidence for 
widespread absorption. For radio-quiet quasars (RQQs), a photon index of $\Gamma\approx$1.9--2.0 is 
obtained (e.g., Vignali et al. 2005; Shemmer et al. 2005), also at $z>5$ (Shemmer et al. 2006a), 
while for ``moderate'' radio-loud quasars (RLQs) and blazars, $\Gamma\approx$1.7 and 
$\Gamma\approx$1.5 are obtained (Lopez et al. 2006), respectively. 
\smallskip\par\noindent$\bullet$ 
The comparison with the lower redshift (luminosity) Palomar-Green quasars observed by 
XMM-{\it Newton} (Piconcelli et al. 2005) indicates that the photon index does not vary 
significantly with redshift and luminosity, but seems to depend primarily on the accretion rate 
(i.e., steeper X-ray slopes 
are associated with higher Eddington ratio sources; Shemmer et al. 2006b).  
\smallskip\par\noindent$\bullet$ 
Following X-ray studies of early '80 and '90, the relation between X-ray and longer wavelength 
emission has been investigated by means of the point-to-point spectral slope between 2500~\AA\ and 
2~keV in the source rest frame ($\alpha_{\rm ox}$). 
Any changes in the accretion rate over cosmic time might lead to changes in the fraction of 
total power emitted as X-rays. Using 333 AGN at $z$$\approx$0--6.3 (88\% X-ray detections), 
Steffen et al. (2006) confirmed that $log~L_{2500~\mbox{\scriptsize \rm \AA}}$ correlates with 
$log~L_{\rm 2~keV}$ with an index $<1$, and $\alpha_{\rm ox}$ depends upon 
$log~L_{2500~\mbox{\scriptsize \rm \AA}}$ (with the slope perhaps depending on 
$L_{2500~\mbox{\scriptsize \rm \AA}}$).

%\section{What's next?}
The research field related to $z>4$ AGN still offers plenty of opportunities. 
In particular, the detection of X-ray variability in some $z>4$ quasars over time scales of 
month-year (Shemmer et al. 2005) needs further investigations to check the possibility that 
quasars are more variable in the early Universe. Furthermore, detailed X-ray spectra of 
$z>4$ RLQs filling the observational gap between ``moderate'' RLQs and blazars are still needed, 
as well as studies of ``peculiar'' quasars and faint AGN population at the highest redshifts.

%\acknowledgements 
%CV acknowledges support from ASI--INAF I/023/05/0. 

%%% THE BIBLIOGRAPHY
%%%
%%% CONSULT SECTION 3 OF "INSTRUCTIONS FOR AUTHORS" FOR HOW TO USE NATBIB.
%%% AUTHORS ARE ENCOURAGED TO USE EITHER THE "THEBIBLIOGRAPY" ENVIRONMENT
%%% BY UNCOMMENTING (DELETING THE "%" SYMBOL) THE COMMANDS BELOW, OR BY
%%% USING THE BIBTEX ENVIRONMENT. TO FIND OUT WHICH IS APPLICABLE TO YOUR
%%% CONTRIBUTION, CONSULT THE VOLUME EDITORS FOR YOUR PROCEEDINGS.
%%%

\end{document}